\documentclass[structabstract]{aa}  

\usepackage{natbib}
\bibpunct{(}{)}{;}{a}{}{,} 
\usepackage{graphicx}
\usepackage{color}

\newcommand{\Av}{A$_\mathrm{V}$}

\begin{document}

\title{
Direct evidence of dust growth in L183 from MIR light scattering
      }


\author{
 {J. Steinacker}\inst{1,2}
\and
 {L. Pagani}\inst{1}
\and
 {A. Bacmann}\inst{3}
\and
{S. Guieu}\inst{4}
       }
\institute{
LERMA \& UMR 8112 du CNRS,
 Observatoire de Paris,
 61 Av. de l'Observatoire, 75014 Paris, France\\
\email{stein@mpia.de}
\and
 Max-Planck-Institut f\"ur Astronomie,
 K\"onigstuhl 17, D-69117 Heidelberg, Germany\\
\and
 Laboratoire d'Astrophysique,
 Observatoire de Grenoble, BP 53,
 F-38041 Grenoble, Cedex 9, France\\
\and
 Spitzer Science Center,
 1200 E California Blvd,
 Mail Stop 220-6,
 Pasadena CA 91125, USA\\
          }
\date{Received 07/07/2009; accepted 30/11/2009}

\authorrunning{Steinacker et al.}
\titlerunning{Direct evidence of dust growth}

\abstract
 {
Theoretical arguments suggest that dust grains should grow in the dense 
cold parts of molecular clouds. 
Evidence of larger grains has so far been gathered in near/mid infrared 
extinction and millimeter observations.
Interpreting the data is, however, aggravated by the complex 
interplay of density and dust properties (as well as temperature for 
thermal emission).
 }
 {
Direct evidence of larger particles can be derived from scattered
mid-infrared (MIR) radiation from a molecular cloud observed in a 
spectral range where little or
no emission from polycyclic aromatic
hydrocarbons (PAHs) is expected.
 }
 {
We present new Spitzer data of L183
in bands that are sensitive and insensitive to PAHs.
The visual extinction \Av~map derived in a former paper was fitted by a series of 3D
Gaussian distributions.
For different dust models, we calculate the scattered MIR
radiation
images of structures that agree with the \Av~map 
and compare them to the Spitzer data.
 }
 {
The Spitzer data of L183 show emission in the 3.6 and 4.5 $\mu$m bands,
while the 5.8 $\mu$m band shows slight absorption.
The emission layer of stochastically heated particles should coincide 
with the layer of strongest scattering of optical interstellar radiation,
which is seen as an outer surface on I band images different from
the emission region seen in the Spitzer images.
Moreover, PAH emission is expected to strongly increase from 4.5 to 5.8 $\mu$m,
which is not seen. Hence, we interpret this 
emission to be MIR cloudshine. 
Scattered light modeling when assuming interstellar medium dust grains 
without growth does not reproduce flux measurable by Spitzer.
In contrast, models with grains growing with density yield
images with a flux and pattern comparable to the Spitzer images
in the bands 3.6, 4.5, and 8.0 $\mu$m.
 }
 {
There is direct evidence of dust grain growth in the 
inner part of L183 from the scattered light MIR images seen by Spitzer.
 }

\maketitle

\section{Introduction} \label{intro}

Cold and dense cores of molecular clouds are commonly considered to be 
a favorable place for the growth of dust particles 
\citep[see, e.g.,][]{1994A&A...291..943O}. 
Evidence that grains are larger than commonly observed
interstellar medium (ISM) grains have been gathered in different
wavelength regions.
The thermal emission of dust grains in the far-infrared (FIR) and 
millimeter (mm) wavelength range shows a change in the index of the 
spectral density distribution. 
Based on simple assumptions about the dust opacity law and the spatial 
and thermal structure, this is interpreted as indirect evidence 
of larger grains
\citep[e.g.][]{2003A&A...398..551S,                                                      2004ApJ...616L..23B,                                                      2006A&A...451..961R,                                                      2006MNRAS.373.1213K,                                                      2008ApJ...684.1228S}.
In contrast, \citet{2008MNRAS.384..755N} can model
the Taurus Molecular Ring without any assumption about grown particles.
In general, disentangling  spatial variations in dust properties,
density, and temperature is a challenge given the many
parameters needed to describe the often clumpy or filamentary
cloud structures. This is why, in view of this ambiguity, we call this 
{\em indirect} evidence.

\citet{2009ApJ...690..496C}, \citet{2009ApJ...693L..81M}, and
\citet{2009ApJ...699.1866C} have presented mid-infrared (MIR) 
extinction curves based on Spitzer and ground-based JHK band observation
of stars behind molecular clouds. 
They find that the extinction curve changes as a function of increasing 
extinction. 
\citet{2009ApJ...693L..81M} investigates the ice features and concludes 
that a process involving ice is responsible for the changing shape of 
the extinction curve.
\citet{2009ApJ...690..496C} find that, with increasing extinction, the 
extinction law is better fitted by models including larger grains.
Similiarly,
for the cores investigated in \citet{2009ApJ...699.1866C}, 
the averaged extinction law from 3.6 to 8 $\mu$m is consistent with
a dust model that includes larger grains.

Extinction measurements at optical to MIR wavelengths
have the advantage that the temperature of the dust grains no longer 
influences the results. The disadvantages are
(i) the method requires a radiation source like scattered light,
diffuse emission from small grains or molecules, or background stellar 
or extragalactic emission;
(ii) the dust extinction properties enter exponentially in the 
observed extinction of the external radiation so that the source flux, 
if stellar or extragalactic, must be strong enough to be detectable.


Aside from the thermal emission of the grains and their absorption, they also efficiently scatter radiation.
The exploration of the physical properties and spatial distribution of 
dust particles via the observation of their scattered light was initiated long ago. References
to early papers about this topic can be found, e.g., in 
\cite{1975A&A....42..357S}.
Common ISM grains with sizes around 0.1 $\mu$m
efficiently scatter optical and near-infrared (NIR) radiation so that images observed at 
these wavelengths can be used to study the spatial structure of the dust, as well
as its properties.
This method has widely been used in star formation research, e.g. to 
analyze the spatial 3D structure of circumstellar disk candidates 
\citep[see, e.g., ][]{2006A&A...456.1013S}.
\citet{2003AJ....125.1407N} interpret the surface brightness seen in 
J, H, and Ks images of the Lupus 3 dark cloud as starlight scattered by
dust. 
\citet{2006ApJ...636L.105F} propose to use this ``cloudshine'' as a 
general tracer of the column density of dense clouds and presented 
NIR
scattered light images of the Perseus molecular cloud complex.
\citet{2006ApJ...636L.101P} present a corresponding method and test it
by calculating NIR scattered light images from the results of 
magnetohydrodynamic turbulence simulations and by comparing the resulting
column density distributions with the original simulation data. They
argue that the method can be applied to filamentary cloud structures being
illuminated by a homogenous radiation field in the range of visual extinction 
between 1 and 20 mag.
The NIR cloud shine of a Corona Australis cloud filament was investigated
in this way by \citet{2008A&A...480..445J}.

It seems promising to observe scattered light from cores at 
longer wavelengths, because the lower extinction allows the inner
parts to be traced. However, the scattering cross section of the ISM dust particles drops with
wavelength once the wavelength is much larger than the particle size. 
Scattered light from cores at MIR wavelengths therefore has not been
reported so far and has been overlooked until now despite submillimeter 
observations often being explained by grain growth to sizes for which MIR 
scattering may become visible.
Additionally, polycyclic aromatic hydrocarbon (PAH) emission starts to 
dominate the emission pattern, at most
MIR wavelengths, preventing the identification of scattered light.

It turns out that indeed
the scattered light image analysis at MIR wavelengths
opens up a {\em window to trace bigger
grains}:
(i) they scatter radiation efficiently at wavelengths for which the
smaller grains already enter the regime of Rayleigh scattering where
absorption dominates scattering;
(ii) the optical depth is reduced compared to the depth at NIR wavelengths 
so that denser regions can be traced with a better chance of seeing grown 
particles.

Observations with the Spitzer telescope allow us to trace the relevant
wavelengths and to check the existence of possible PAH emission by using a
combination of filters, some including strong emission features and
some excluding all.
Nearby starless low- to mean-mass clouds are potential candidates 
for observing such an MIR cloud shine.

{L183} (also known as {L134N}) is such a starless dark 
cloud high above the Galactic plane (36$^o$), hence close to us 
\citep[110 pc,][]{1989A&A...223..313F}. 
Its mass has been estimated to $\sim$80\ M$_{\odot}$  
\citep[][hereafter Paper I]{2004A&A...417..605P}. 
It contains an elongated ridge of dense material with two prestellar 
cores and the main prestellar core has a peak extinction of 
\Av\ $\approx$ 150 mag with a dust temperature close to 7 K (Paper I). 
It was subsequently shown that the gas is 
thermalized with the dust in the core \citep{2007A&A...467..179P}.

The Spitzer observations of L183 are summarized in Sect.~\ref{data}.
In Sect.~\ref{AV}, we fit the \Av~map presented in Paper I with a
series of 3D density basis functions.
The results of a scattered light image modeling are presented
in Sect.~\ref{scattered} based on a dust growth model
and compared to the different Spitzer images of L183.
In Sect.~\ref{conclusions}, we summarize our findings.

\section{Spitzer observations of L183} \label{data}


To describe different parts of the cloud, we refer to the
\Av{}-map presented in Paper I
(see their Fig.~1 for a detailed comparison),
although it is a projected view on the spatial cloud structure.
Throughout this paper, we use the term 
{\em outer cloud region} for the parts of the L183 cloud where \Av $<$ 15, 
(banana-shaped) {\em central part} for 15 $<$ \Av $<$ 90, and 
{\em central core} for the densest region with \Av $>$ 90,
respectively.

The Spitzer IRAC (3.6 to 8 $\mu$m, \citealt{Fazio-2004}) observations 
of L183 were obtained as part of the  Cycle~1 GTO program 94 
(PI: Charles Lawrence). 
The observations were broken into two epochs, the astronomical 
observation requests (AORs) are both centered at 15h54m15.00s~-2d55m0.0s,
they were obtained on 23 and 25 of August 2005, and the
AORKEYs are 4921600 and 4921856 in chronological order. 
Each AOR was constructed with the same strategy: 
at each map step, three dither positions 
were observed, each with high-dynamic-range exposures
of 1.2 and 30 seconds.

To build the full mosaics of L183, we started
with the Spitzer Science Center (SSC) pipeline-produced, basic
calibrated data (BCDs), version S14.0. 
We ran the IRAC artifact mitigation code 
written by S.~Carey and available on the SSC website. 
We constructed a mosaic for each epoch from the corrected BCDs 
using the SSC mosaicking and point-source extraction (MOPEX) software
\citep{Makovoz-2005}, with a
pixel size of 1.22"\,px$^{-1}$, very close to the native pixel 
scale.

\begin{figure*}
\includegraphics[width=8cm,angle=-90]{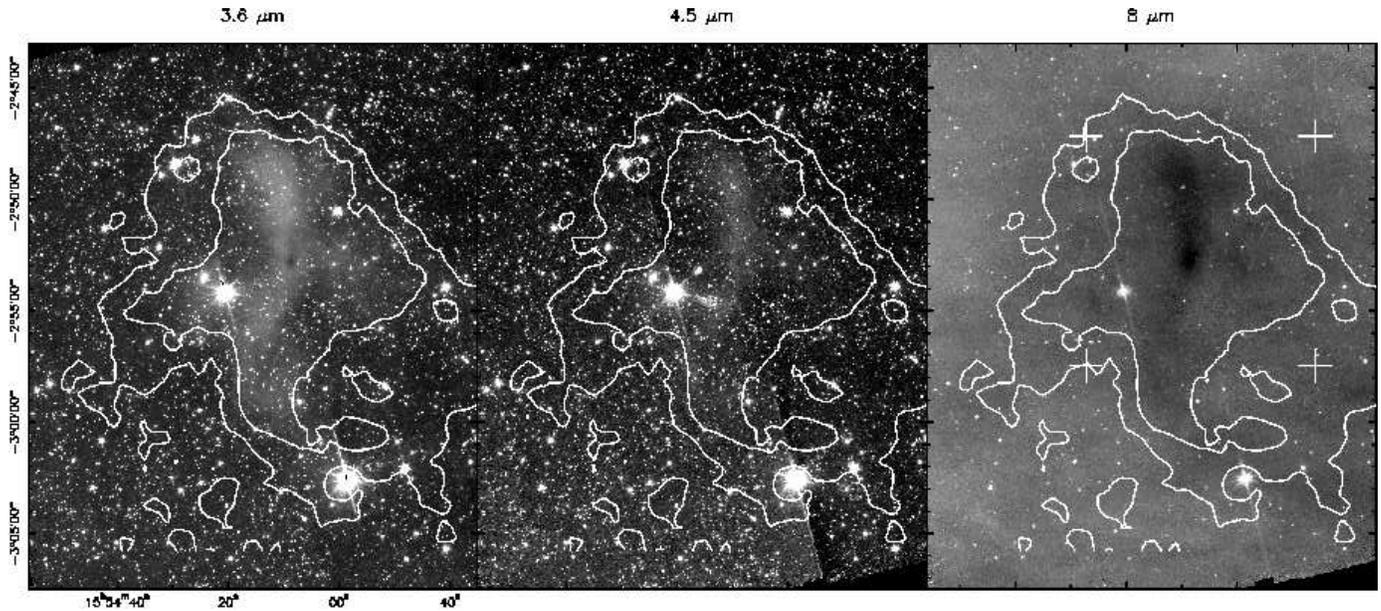}
\caption{IRAC 3.6, 4.5, and 8\,$\mu$m long exposure mosaics. Superposed on each image are the \Av\ = 5 \& 10 mag contours from the \Av\ map of Paper I. 
The white crosses in
the 8 $\mu$m image mark the region that has been modeled in this paper.}
\label{IRACmosaics}
\end{figure*}

The inner parts of 3.6, 4.5, and 8\,$\mu$m mosaics are
shown in Fig.~\ref{IRACmosaics}.
One can clearly see a banana-shaped structure (see also 
Sect.~\ref{scattimages}) visible in emission at 3.6 and 4.5\,$\mu$m and in
absorption at 8\,$\mu$m. 
The spatial structure resembles well the shape of the densest region
of the cloud complex visible in the \Av~map presented in Paper I.
The 5.8 $\mu$m image shows horizontal stripes and a strong vertical
gradient in signal-to-noise ratio (S/N) so that only horizontal pieces of the upper part of 
the image contain reliable information on the flux. For this reason,
we do not show the image here, and do not include it in the image
modeling. The upper part of the central banana-shaped structures is
seen in absorption in the reliable stripes.

\begin{figure}
\includegraphics[width=5.5cm,angle=-90]{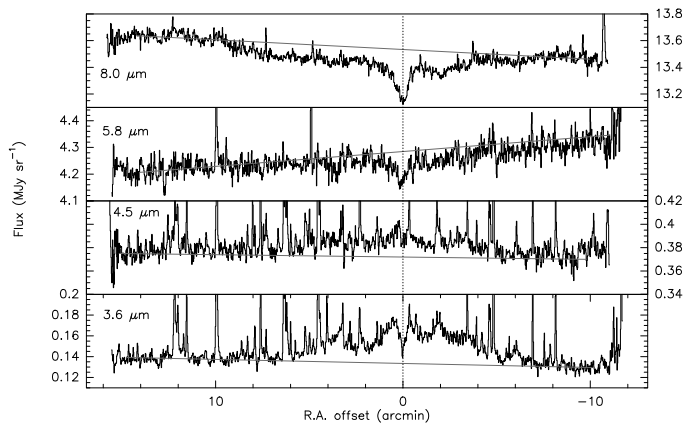}
\caption{Cuts from the IRAC maps (Fig.~\ref{IRACmosaics}) at constant 
declination (-2$^o$52'48") through the dust peak in all 4 bands. 
The dotted line indicates the position of the dust peak. 
Excess emission is seen at 3.6 and 4.5 $\mu$m while absorption 
dominates for the 5.8 and 8.0 $\mu$m ranges.}
\label{horcuts}
\end{figure}

To show the magnitude of the emission and absorption features, we 
present in Fig.~\ref{horcuts} cuts of the IRAC maps through the
central core (declination -2$^o$52'48") seen in absorption on all images.
As the cut lies in a stripe of the 5.8 $\mu$m image with an acceptable
S/N, we also include the flux in this band.
The gray thin lines indicate an averaged external flux. From the
cuts, the increase by almost two orders of magnitude in the external
flux is visible when changing from 3.6 to 8 $\mu$m.
 
The thermal emission of dust particles which are large enough to be
considered as black body emitters cannot produce a measurable flux
given the low temperature of a few ten K \citep{2004A&A...417..605P}. 
Even if all dust particles in L183 would have a temperature of 40 K,
the specific flux density measured on Earth at
$\lambda=8\ \mu$m would be $<7\times
10^{-29}$ Jy/sr (assuming 0.1 $\mu$m-sized grains and a gas-to-dust mass
ratio of 100), and lower for the shorter IRAC band wavelengths. 

\begin{figure}
\includegraphics[width=\hsize]{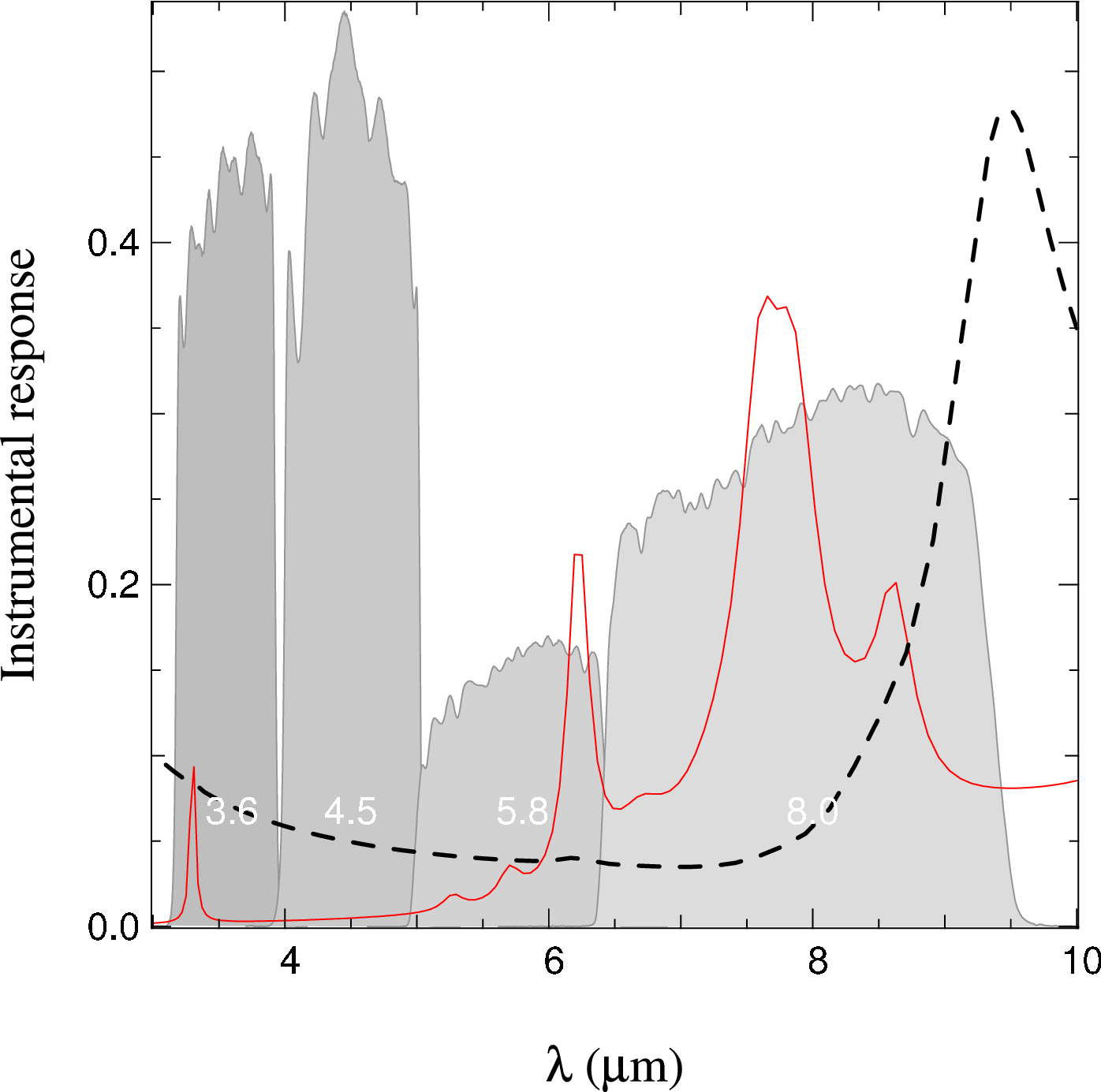}
\caption{The gray areas are the instrumental response of each IRAC 
filters as a function of the wavelength. 
The light gray (red in the electronic version) line is a model of infrared emission from dust grains by 
\citet{Draine-2007} (q\_PAH=4.58\% U=1). 
The black dashed line is a model of the dust extinction
from \citet[see][]{Draine-2003a,Draine-2003b,Draine-2003c}. 
The models are unitless, and have been scaled to fit in the figure.} 
\label{IRACfilters}
\end{figure}

The emission in the Spitzer wavelength range could also be caused by
smaller particles that are stochastically heated like PAHs.
Figure \ref{IRACfilters} illustrates the presence of PAH emission features
in IRAC bands by showing the instrumental response of IRAC (gray areas) for the 
different filters as a function of wavelength. 
The transmission of
radiation through a layer of dust is plotted as a black dashed line 
\citep[see][]{Draine-2003a,Draine-2003b,Draine-2003c}.
The light-gray (red in the electronic version) line shows the expected 
location of PAH features \citep{Draine-2007}.

Explaining the observed emission in the 3.6 and 4.5 $\mu$m bands as the
emission of stochastically heated particles (PAHs or small dust grains)
is difficult for two reasons.
First, PAHs and small grains need an exciting UV to optical radiation field.
As dust particles have the same order of magnitude in absorption and
scattering cross section at these wavelengths, the scattered light images
at optical wavelength give a hint of the morphology of the layer where most of the outer
exciting UV to optical radiation is absorbed.
As seen, e.g, in Fig.~4d of \citet{2002A&A...382..583J}, this layer
resembles the shape of the \Av=1-layer in the \Av~map closely (Paper I).
In contrast, the emission pattern resembles the shape of the inner banana-like
structure while no emission can be seen with the shape of the outer
layer. Even if holes would make it possible for exciting radiation to reach
inner parts, a substantial part of the PAH emission should come from
the outer layer which is not what is seen.

For completeness, we also investigated the possibility of holes.
We performed detailed 3D ray-tracing
for a large sample of structures that are consistent with the
column density information
contained in the \Av~map.
The results are presented in Appendix~\ref{structure}.
Based on
3D spatial structure variations
and the analysis of $\tau$-iso-surfaces, we conclude that only
a small fraction of all spatial structure models would
allow for PAH emission near the central absorption
peak. 
It therefore seems unlikely that the cloud complex has one
of these rare configurations with a hole deep enough to let the
interstellar radiation field (ISRF) reach the central
part so that PAHs could be excited to produce the observed
inner emission.
Moreover, the illumination through a single hole would not produce a 
smooth, even illumination of the entire inner structure as seen in the
3.6 $\mu$m image.

Second, based on an integration of the gray/red curve in Fig.~\ref{IRACfilters}, the PAH emission in the 5.8 $\mu$m band should be a factor of
about 10
stronger than the emission in the 4.5 $\mu$m band. The improved
dust transmission in the 5.8 $\mu$m band will amplify this effect.
In turn, the external radiation at 5.8 $\mu$m is about a factor of 10 stronger
than at 4.5 $\mu$m. 
Based on \citet{2006A&A...453..969F}, we estimated the flux one would expect if the excess flux at 4.5 $\mu$m was due 
to PAHs. We found that, in all reasonable cases, the 5.8 $\mu$m flux should be
seen in emission (about 1 to 2 orders of magnitude stronger than at 
4.5 $\mu$m) and not in absorption as presently observed (see cuts in Fig.~\ref{horcuts}).

These two arguments suggest that emission from stochastically heated 
grains likely fails as an explanation of the emission structure.
We therefore explore in this paper whether the emission seen in
the 3.6 and 4.5 $\mu$m filters can be interpreted as stellar light scattered
by the dust particles in the cloud gas, and what dust properties 
would be needed to qualitatively reproduce the images.


 \section{\Av~map modeling} \label{AV}

Extinction maps average the 3D density distribution
of the molecular cloud over the line of sight (LoS). This adds an
important source of ambiguity to the interpretation of the extinction map:
each observed extinction value in the plane of sky agrees with
a whole range of density distributions and dust properties.
As a result, another source of information is needed to remove this ambiguity.
This can be, e.g., a measurement at FIR/mm wavelengths. 
Multi-wavelength modeling of images from molecular clouds and cloud cores
incorporating the hidden
third dimension has not been used widely so far, but is possible in the
mean time
\citep[see, e.g.,][]{2005A&A...434..167S}.
Another source of information are scattered light images. They depend
on the 3D spatial structure of the cloud, on the dust properties, and the
external 3D radiation field.

In this work, we combine the information about L183
from the extinction map with the scattered light images obtained by
Spitzer.
To obtain a 3D dust density data cube required for the scattering calculations,
we fit the available \Av~map of L183 (Paper I) with a series of density
basis functions flexible enough to deal with the complex 
spatial structure of the cloud. The fit of the 2D \Av~map is done
by determining the optical depth along the LoS through a 3D density
structure described by the series. This means that parameters describing
the spatial structure along the line of sight will not be determined by this fit.
Instead, a library of possible spatial structures is created and used for the 
modeling.
Details of the technique to fit the \Av~map are described in
\citet{2005A&A...434..167S}, but here we just summarize the basic approach
and refer readers to that paper for details.
The dust properties are fixed for this determination to $A=$0.1 $\mu$m
silicate spheres, but in later sections we re-interpret the obtained
optical depths using other grain sizes.

The underlying 3D density structure is described by a series of 100
3D Gaussian density clumps with their main axes aligned with the
Cartesian coordinate axes. The position and extent in the plane of sky of each Gaussian (4 parameters) and their normalization factor (5th parameter) are
optimized with a special simulated annealing search tool
\citep{1994A&A...287..493T}. The high number of Gaussians is necessary
because the \Av~map is very clumpy with many local maxima and structure on small 
and large scales. Tests with lower numbers of Gaussians
could not reproduce the main features of the map adequately. 
It must be stressed, however, that the original map in Paper I is 
incomplete (lower limit to the extinction toward the northern prestellar 
core) and that it probably suffers from aliasing 
in the intermediate opacity 
range (\Av = 20 -- 40 mag) because there are few background objects.

\begin{figure}
\centering
\includegraphics[width=9cm]{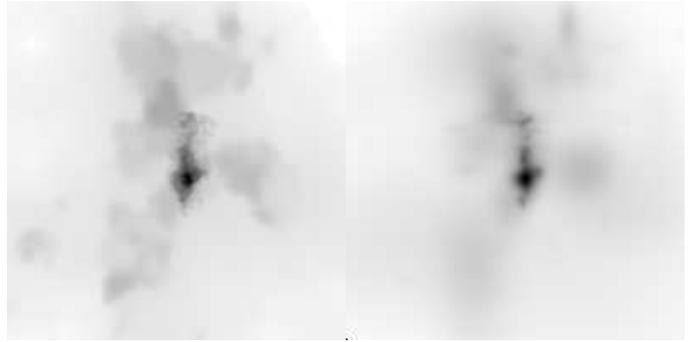}
\caption{
Inner 66000 AU of the \Av~map from Paper I (left) and the fitted map using 
3D Gaussian clumps (right).
        }
\label{AVmapandfit}
\end{figure}

In Fig.~\ref{AVmapandfit}, we show the inner 66000 AU of the \Av~map 
(left) and the corresponding fitted map using Gaussian clumps (right).
\begin{figure}
\centering
\includegraphics[width=9cm]{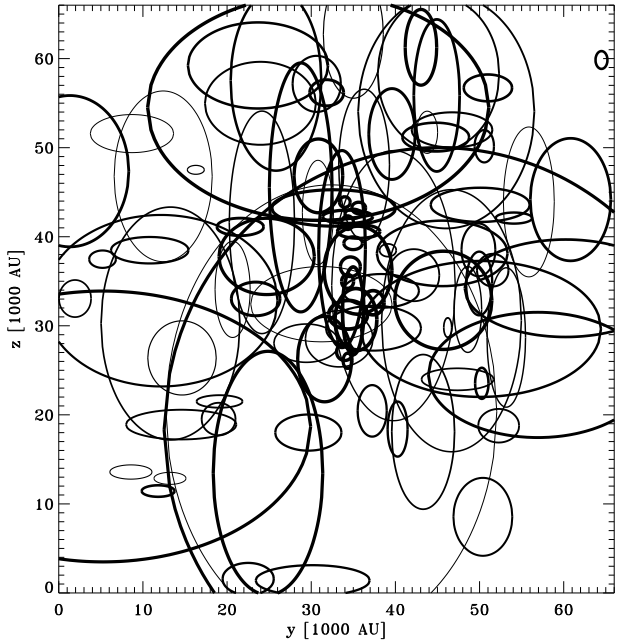}
\caption{
Position and main axes of the Gaussian basis functions used to fit the
inner part of the \Av~map. The coordinates $x$ and $y$ are in the plane
of sky.
        }
\label{Gaussians}
\end{figure}
\begin{figure*}
\centering
\includegraphics[width=18.7cm]{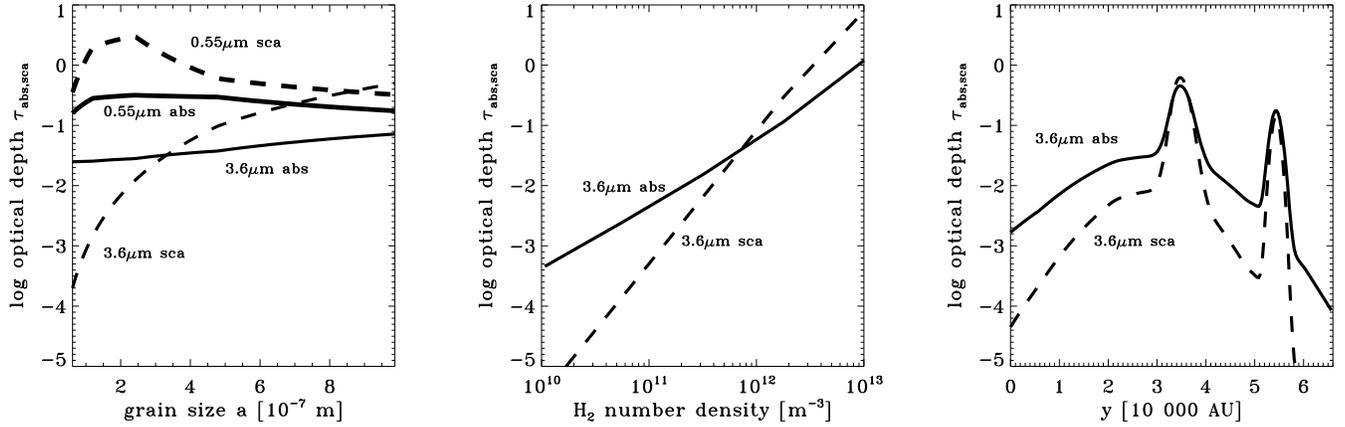}
\caption{
Left:
Local optical depth in a cell of the 3D density structure (constant 
density)
as a function of the grain size (assuming dust mass conservation).
The solid (dashed) line refers to absorption (scattering), and the
thick (thin) line to a wavelength of
0.55 $\mu$m (3.6 $\mu$m).
Middle: Optical depth for dust following Eq.~(\ref{grow}) as a function of
 the cell gas number density for
absorption and scattering at $\lambda=$3.6 $\mu$m
($a_0=6\times 10^{-8}$ m, $\alpha=0.4$, $n_{H_2;0}=10^{10}$ m$^{-3}$).
Right: Local absorption and scattering optical depth along a typical ray crossing one of the 10,000 3D density data cubes that fit the \Av\ map.
        }
\label{tautau}
\end{figure*}
The underlying Gaussian distribution is sketched in Fig.~\ref{Gaussians}
by showing ellipses with the main axes of the Gaussians at their
corresponding position in the plane of sky. 
The line thickness is a measure of the mass in
the different clumps.
The figure shows that the clumpy structure of the \Av~map requires a series
of basis functions in the inner 10000 AU along the x-axis
to reproduce the bananashape of the cloud center. In turn, some of the detailed
spatial structures (e.g. in the lower left part of the \Av~map) are represented by
a low number of basis functions, which indicates that the chosen number of 100
is still not enough to describe all features of the map.

The fit of the \Av~map provides us with a column density structure,
but the extent and position of the Gaussians along the line of sight
are still unknown. 
To provide a basis for the fits of the scattered light images, we have created 
a library of 10000 3D density structures where the spatial structure parameters
along the LoS were chosen randomly given the outer physical borders and the
constraint to agree with the fit of the \Av~map as shown in 
Fig.~\ref{AVmapandfit}.

\section{Scattered light image modeling using a dust growth model} 
   \label{scattered}
  \subsection{Dust model} \label{dust}

It is expected that in the cold core environment,
the particles
grow to fluffy aggregates, 
vary in their chemical composition and follow an entire
size distribution \citep[see, e.g.,][]{1993A&A...280..617O}.
While there are no direct measurements of the dust grain
size distribution in
dense cores, model calculations and laboratory experiments at least
can provide rough estimates of the mean grain size as a function of
the gas number density in one free-fall time 
\citep{2005A&A...436..933F}.
Therefore, we assume that the grains
have properties approximately described by spherical homogeneous
particles of one size and chemical composition.
The radius of the grains $a$ is assumed as a simple powerlaw of the 
$H_2$ number density $n_{H_2}$ with the spectral index $\alpha$
\begin{equation}
a=\left\{ 
    \begin{array}{ll}
        a_0\left(\frac{n_{H_2}}{n_{H_2;0}}\right)^\alpha & n_{H_2} > n_{H_2;0}\\
        a_0 & n_{H_2} \leq n_{H_2;0}\\
    \end{array} 
\right.
\label{grow}
\end{equation}
so that the grains in regions below the threshold density $n_{H_2;0}$
have the size $a_0$.
The calculations leading to the best-fitting
results described in this section were
performed with varying values for $a_0$, $n_{H_2;0}$, and $\alpha$. 
\citet{1997AdSpR..20.1595P}, e.g., found a growth index of $\alpha=0.25$ 
\citep[see][their Fig.~3, plotting the dependence in
double--logarithmic scale]{2005A&A...436..933F}.

In Sect.~\ref{AV}, we modeled 
the local optical depth for extinction $\tau_{ext}(\lambda)$ for $\lambda=0.55\ \mu$m
toward the observer for each cell of our spatial grid.
For the radiative transfer modeling of the Spitzer scattered light images, 
we need to calculate the optical depth for absorption and scattering at 
the Spitzer band wavelengths in each cell. If we know the grain size
and the number density of these grains, $n$, in each cell, 
we can use existing opacity tables to find the cross sections $\sigma_{abs/sca}(\lambda,a)$ and calculate
the optical depths. To use Eq.~(\ref{grow}), this requires finding the gas density
from the derived local extinction $\tau_{ext}(0.55\ \mu$m$)$.

The optical depth
across a Cartesian grid cell with the edge length $s$ is
\begin{equation}
\tau_{ext}(0.55\mu\mathrm{m})
=
 \sigma_{ext}(0.55\mu\mathrm{m},A)\ n_0\ s,
\end{equation}
if all particles have the size $A$ and $n_0=n(A)$.
Assuming an internal density of each homogeneous grain $\rho_d$, the mass
of a grain is $m_d=4\pi\ A^3\ \rho_d/3$, and the total dust mass in the
cell volume is $n_0\ m_d\ s^3$. 
For a given mass gas-to-dust ratio $R_{g/d}$, this defines
the H$_2$ mass $n_{H_2}\ m_{H_2}\ s^3$, hence 
\begin{equation}
n_{H_2}
=
\frac{4\pi A^3\ \rho_d}{3m_{H_2}}\ 
R_{g/d}\ 
\frac{\tau_{ext}(0.55\mu{}m)}{\sigma_{ext}(0.55\mu{}m,A)}\ 
\frac{1}{s}.
\end{equation}
This allows us to calculate the H$_2$ densities for all cells and
the corresponding grain size using
Eq.~(\ref{grow}).
The corresponding number density $n(a)=n_0\ A^3\ a^{-3}$
results from the conservation of dust mass in the cell.
Using the
\citet{1984ApJ...285...89D} opacities, we can then calculate the
corresponding local MIR optical depth 
$\tau=\sigma_{abs/sca}(3.6\mu{}m,a)\ n(a)\ s$ for the absorption
and scattering needed for the radiative transfer modeling of the
scattered light images.

The effect of grain growth on optical depth due to
absorption and scattering is illustrated in Fig.~\ref{tautau}.
The left plot shows the optical depth for absorption and scattering
through a typical cell of the 3D density structure with constant density
for $\lambda=$0.55 $\mu$m and 3.6 $\mu$m as a function of the
grain size. The grain numbers have been chosen to conserve the
dust mass in the cell.
For optical wavelengths, the change in the ratio of 
$\sigma_{abs}$ and $\sigma_{sca}$ 
is less than
a factor of 4 over a tenfold increase in size.
In contrast, at 3.6 $\mu$m, absorption becomes more efficient by a factor of
3, while scattering is amplified by a factor of 300.
Even a moderate increase of grain size by a factor of 2 will result in
about 10 times more efficient scattering. 
Correspondingly, the MIR wavelength range allows to see larger grains
in scattered light as scattering becomes efficient again for higher
gas number densities above about $n_{H_2;0}$ if the optical depth for
absorption allows the scattered light to proceed toward the observer. 
The dependence of the optical depth for absorption and scattering
on the gas number
density when using Eq.~(\ref{grow})
is illustrated in the middle figure.
In the densest regions,
the optical depth for scattering is larger than for 
absorption at 3.6 $\mu$m.
The right plot gives the optical depths for a typical ray cutting 
an entire density cube consistent with
the \Av~map of L183 and including grain growth. 
In this figure we see that while we cross clumps of high density, scattering becomes 
stronger, so that {\em scattered light should be seen near the densest
regions} in contrast to
NIR light, for which more scattering 
is seen in the outer parts.
Whether scattered light indeed reaches the observer from the inner
parts is a complex interplay 
of background radiation, optical depth, and forward scattering, and
can only be answered by running a radiative transfer code.
\begin{figure}
\centering
\includegraphics[width=9cm]{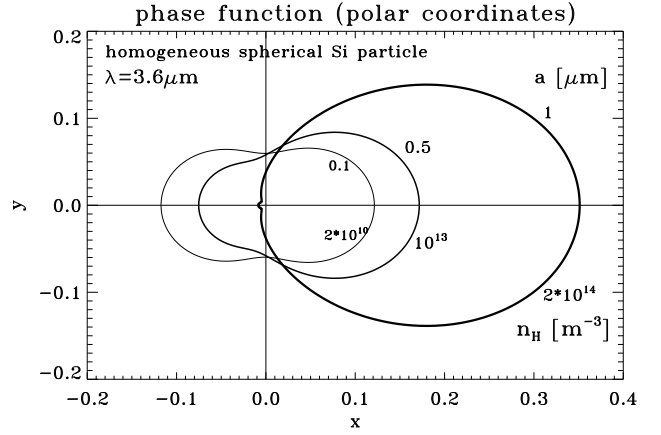}
\caption{
Phase function in polar coordinates of a homogeneous spherical
silicate particle for different particle sizes $a$ at a wavelength of
3.6 $\mu$m. The H$_2$ number densities corresponding to each $a$ as defined
in
Eq.~(\ref{grow}) are given in the lower right part.
        }
\label{phasef}
\end{figure}

Aside from the change in the scattering efficiency due to the dust
particle growth, the scattering phase function describing how
radiation is scattered from one direction to another also changes.
Figure \ref{phasef} shows a typical phase function in polar coordinates
of a homogeneous spherical silicate particle at a wavelength of
3.6 $\mu$m \citep{2003A&A...401..405S}
placed at the origin.
While small ISM particles scatter almost isotropically
at this wavelength, the scattering becomes peaked in the forward direction
for larger grains. 

This forward beaming can alter or even dominate the appearance of the 
scattering images.
For the outer region of the core, 
the interstellar radiation field (ISRF)
reaches the dust from many directions and is scattered almost
isotropically.
For the central region, mainly forward scattering dominates because of
grown particles.
The interstellar radiation from the direction of the observer will not be scattered in a backward
direction. Radiation hitting the core in the plane of the sky also has a low efficiency
to be scattered by 90$^o$. Only radiation from the background is
scattered efficiently in a forward direction toward the observer.
But this radiation has to pass the central parts of the cloud first and
will be seen only if the optical depth is still so little that 
emerging flux can be detected.
For Spitzer, this might actually limit the possibilities of observing the 
growth of grains in the regions of highest density. However, with the sensitivity 
of the mid-infrared instrument onboard the James Webb Space Telescope, 
the wavelength range $>$ 5 $\mu$m will become important for tracing larger
grains.

  \subsection{Theoretical cloud shine images of L183 and dust-modeling
              results} 
   \label{scattimages}

\begin{figure*}
\centering
\includegraphics[width=18.3cm]{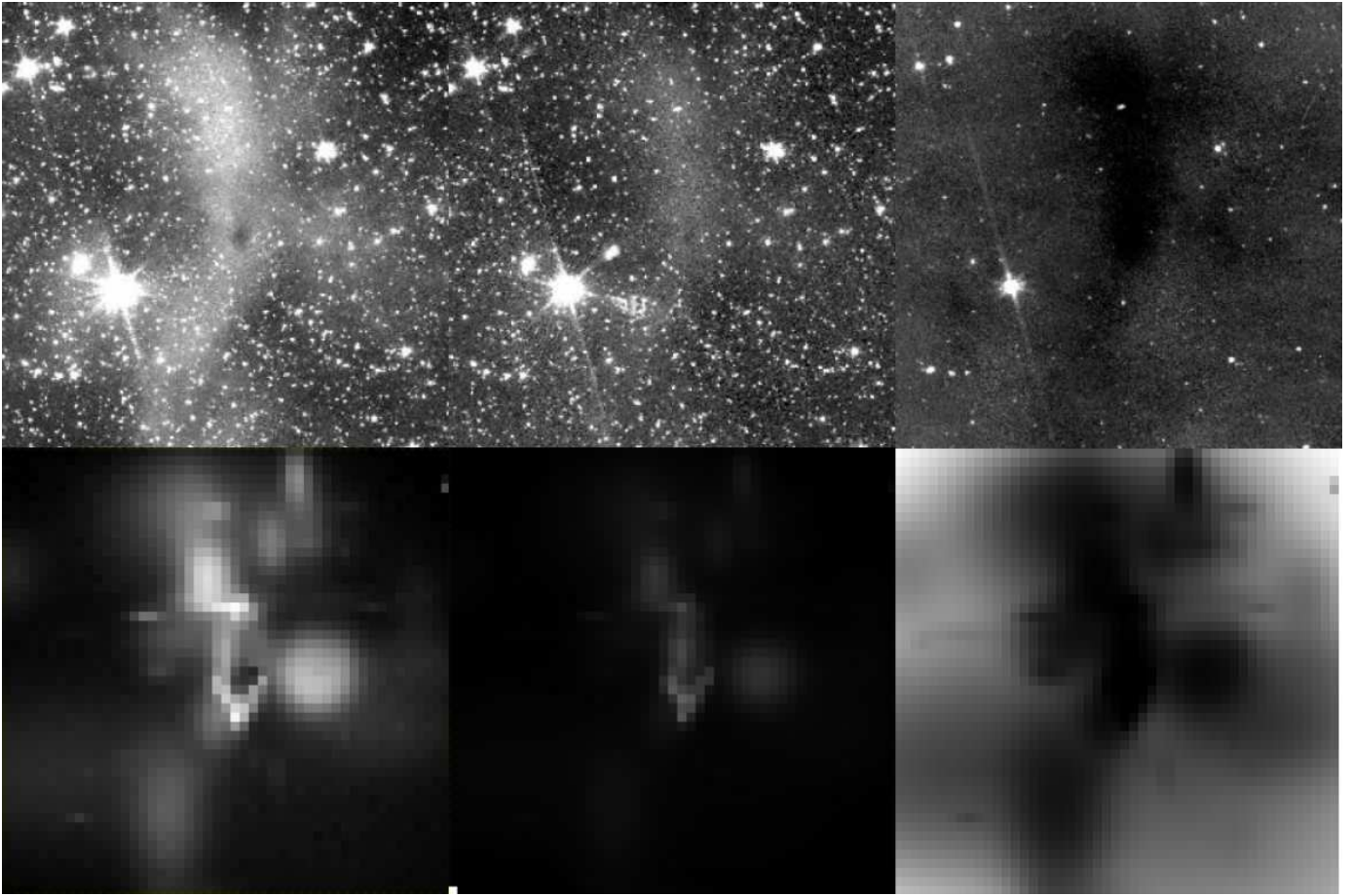}
\caption{
Comparison of the three Spitzer images at 3.6, 4.5, and 8 $\mu$m
of the inner 66000 AU of L183
(top) with scattered light models 
based on grains growing as a function of
density (bottom).
The underlying 3D structure of the model images
is consistent with the measured \Av~map.
The general pattern of the modeled
diffuse emission is similiar while, the flux is about a factor of 2 lower
in the model.
        }
\label{model2}
\end{figure*}
The presentation of the image modeling results in this section has two goals:
(i) to demonstrate that scattered light images based on a density 
structure in agreement with the current extinction data of L183 can
reproduce the main features of the observed Spitzer images, and 
(ii) to present evidence that only models with grain growth can 
reproduce the 
appearance of the central banana-shaped region
in contrast to models with ISM grains without growth.

For the first goal, it is sufficient to present one image with a distinct
set of parameters of the density structure model,
the dust data, and the external field that shows overall agreement.
The common approach for the second goal would be an automated fit and
exploration of the parameter space quantifying the agreement on the
basis of a $\chi^2$-minimization as we used it for the \Av{}-fit.
An automated fitting procedure varying the many free 
parameters is prohibitive, though as
radiative transfer calculations applied to a clumpy 3D structure like 
L183 including scattering are time-consuming.
For the images shown in this section, we used 50$^3$ spatial
grid cells and
a discretization of the direction space in 648 directions
leading to a 
intensity solution vector with 8.1$\times$ 10$^7$ values.
An automated fitting procedure would need to repeat the calculation of
this vector at least several ten thousand times to optimize the density
structure.

Therefore, we present here the results of a nonautomated
exploration based on the
10000 spatial structures derived from the \Av~map fitting as a 
representation of possible density distributions of the cloud.
The quality of the fit was decided by eye: 
reproduction of the central banana-shaped region and its maximum
in the northern part,
the radiation depression in the central core, 
the overall agreement with the flux level observed with Spitzer,
and the shape and location of the weak emission patterns in the outer 
region.

Using the
ray-tracing version of our 3D radiative transfer code 
\citep{2003A&A...401..405S}, we 
selected spatial structures, which yield images fitting the Spitzer 3.6 $\mu$m
image for varying dust models. 
Then we further changed both the structure and the dust 
parameters by hand to study the influence of each parameter and
to find the best-fitting parameter set. Finally, 
we calculated the images
at 4.5 and 8.0 $\mu$m for the same parameter set.

As an external radiation field, we used the DIRBE data provided by
the legacy archive microwave background data analysis (LAMBDA\footnote{http://lambda.gsfc.nasa.gov}) to take the global anisotropy of the ISRF into account.
Moreover, a nearby star could be another (local) 
source of anisotropic illumination.
Investigating the stars in the vicinity of L183, we found one star that could
be near or in the cloud. In Appendix \ref{star}, we discuss the spectral energy distribution (SED), and the resulting stellar properties. A star consistent with the 
SED and placed near or even in the outer parts of L183 would locally
influence the scattered light image at 3.6 $\mu$m up to distances of 2000 AU from
the star. In all other parts of L183, the ISRF dominates the radiation
field.
Therefore, we left the star out in our consideration, like the other
stars visible in the IRAC images.

Figure \ref{model2} compares the Spitzer images with
the resulting scattered light model images. 
The upper images show the inner 66 000 AU of the cloud at
3.6, 4.5, and 8 $\mu$m as observed by Spitzer (see Fig.~\ref{IRACmosaics}).
Images calculated without growth of the grains were too low in scattered
light flux at 3.6 $\mu$m for various ISM grain sizes, and 
all fluxes were below $10^{-5}$ MJy/sr.
The lower images show the best-fitting model images at 3.6 (left),
4.5 (middle), and 8.0 $\mu$m. The corresponding
parameters of the growth model 
as described by Eq.~(\ref{grow}) are $\alpha$=0.4 for the powerlaw
index, a grain size in the outer region of $a_0=5\times 10^{-8}$\,m, 
and a H$_2$ number density threshold for growth at 
$3\times 10^{10}$ m$^{-3}$.
In the lower left hand image, the overall emission pattern is well-matched
with more emission in the northern part.
The model images show some peaks (e.g. directly below the central core), which
obviously are not present in the Spitzer images. Investigating the underlying
density structure, we found that 
those are attributed to single Gaussian clumps and the
image peaks could not be removed by changing their LoS position or extent.
We relate them to either a remaining error in fitting the \Av~map or
to uncertainties in the \Av~map itself.

The lower middle 4.5 $\mu$m image also shows overall agreement with the
corresponding Spitzer image. The same flux normalization
is used as in the 3.6 $\mu$m image. The upper part of the central 
banana-shaped region is visible, and the flux ratio in the observed and
theoretical images F$_{4.5\mu m/3.6\mu m}$ is about 0.2.
The 3.6 $\mu$m Spitzer image shows the central core in absorption
and also in the 4.5 $\mu$m image this dip is weakly seen (compare with
Fig.~\ref{horcuts}). In the model, the dip is reproduced
at both wavelengths. The actual size of the dip is not reproduced 
because of the
limited size of the computational cells.

For the 8 $\mu$m image, a background flux value was interpolated from
regions of low extinction in the Spitzer image and 
assumed as a constant background.
Therefore, comparing the image
with the Spitzer data has to be done with caution:
bright regions in the lower right image 
correspond to regions where
the background radiation experiences little extinction.
The calculations include scattering (which is very low even in the
presence of larger grains), but do not consider the effect of PAH
emission by the cloud itself. 

As absorption dominates at all densities, the image reproduces
the main features of the \Av~map. It also reveals where the density
structure model was still too coarse: single Gaussian clumps appear
where the Spitzer image shows more complex absorption structures.
The overall shape nevertheless agrees with the observed
data. We also cannot exclude the influence of variations in the
background.

The fitting process revealed a prominent {\em ambiguity} in the 
obtained best-fitting dust growth parameter sets. Comparing the sets, we
found that they are based on the dominance of larger particles to
create scattered light in the MIR. As a result, the images were not
sensitive to the density threshold and the outer particles
size, as long as the outer particles are smaller than 0.1 $\mu$m to
avoid substantial scattered light flux in the outer parts.
This ambiguity is visible in the growth law defined in Eq.~(\ref{grow}):
changing the density threshold by a factor $f$ can be compensated
for by adjusting the size of the outer grains by a factor $f^\alpha$
to get the same shape of the grain growth.
In contrast, changing the spectral index had a strong effect on the 
images. Already for $\alpha=0.3$, the flux values were too low by
a factor of 4.
\begin{figure*}
\centering
\includegraphics[width=18.3cm]{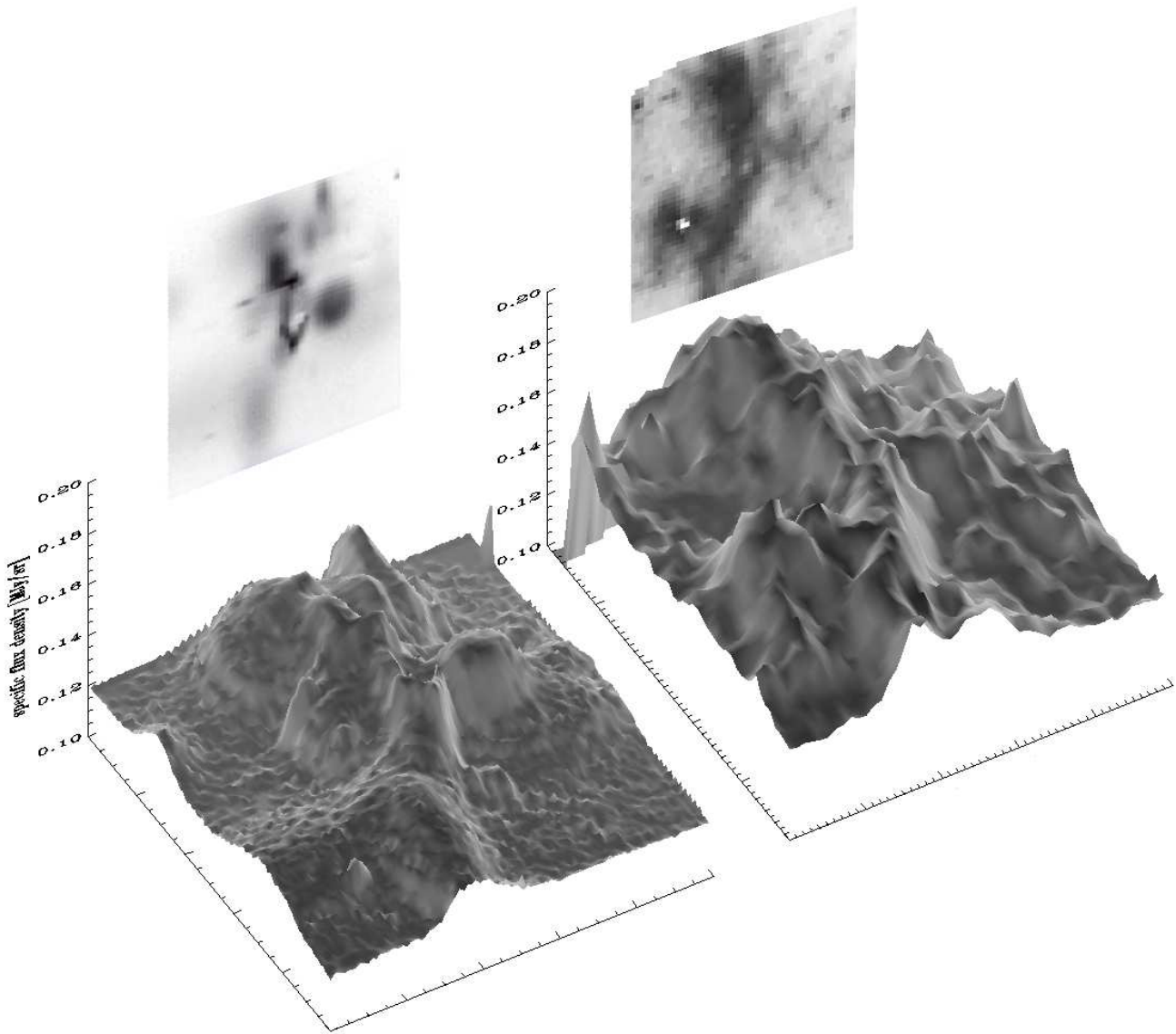}
\caption{
Comparison of the specific flux density (z-axis) of the 
3.6 $\mu$m Spitzer image 
of the inner 66000 AU of L183 (right), processed through a star removal routine,
with the scattered light model image (left).
        }
\label{fig9}
\end{figure*}
 
In general, the flux of the theoretical images is about a factor of 2 lower
than the observed flux. We show the specific flux
density of model and observed image in Fig.~\ref{fig9} both as flat image
and in surface representation to compare the actual values.
As the
stars confused the image substantially, 
we have processed the 3.6 $\mu$m IRAC image through a
routine that removes the stars. The routine considers the flux distribution
of subimages and marks pixels with fluxes in the tail of the
distribution, then interpolates them from the remaining pixels.
The emission peak near the bright star in the lower left part of the
image is an artefact from the spread-out PSF of this star.

While the flux pattern could be reproduced by the model, the higher flux
could not be achieved by changing the grain properties or the spatial
structure.
Within the simple dust model, 
one might consider that higher flux value can be achieved by assuming
a stronger growth of particles with the bigger grains producing more
scattered light. The increased extinction, however, damps this radiation
more efficiently so that the emission of these grains cannot be
observed any more.

The flux will also change if we consider a more sophisticated dust 
model. As an example, if we consider elliptically shaped grains
with perfect grain alignment to a large-scale magnetic field, 
the phase function, hence the
scattered light flux, will increase by a factor of 2 by assuming 
a moderate grain main axes ratio of just 2 
\citep[see, e. g.,][]{1993Ap&SS.204...19V}.

It can be expected that an automated fitting and/or an improved modeling
of the \Av~map would remove most of the obvious peaked artefacts. 
Further improvement can be expected from improving the \Av~map in the 
northern
banana-shaped region
because in the original study (Paper I), that part was not mapped with 
the MIR camera of the Infrared Space Observatory (ISOCAM). 
From mm measurements, the expected extinction reaches \Av\ $\approx$ 
60 to 70 mag (being somewhat insecure for technical problems 
with the bolometer). 
This was beyond the sensitivity of the H-K'\ color-reddening 
method used to map this part to complement the ISOCAM map. 
Lower limits in the range \Av\ = 25--40 mag have been filled in 
instead, which did not induce dense clump fitting in the 3D \Av\ model. 

In a more realistic model, larger grains might be expected
to grow as a second population from the original seed population. So
there is the normal Rayleigh scattering by most particles and some more
efficient scattering of bigger grains. In the simple model in the paper,
all grains are moved away from the Rayleigh regime when growth occurs.

In test runs, we also estimated the limit for the largest grains we
are able to detect with our method in L183 based on our dust model. 
For 3.6 and 4.5 $\mu$m, grains of
about 1 $\mu$m in size make the largest contribution to the scattered
light flux. The flux from larger grains quickly drops because of the 
strong extinction within the dense regions.

An interesting point is the timescale of the growth process related
to the evolution time scale of the core and cloud. If
the clouds are long-lived, grains might grow enough to become a seed
population for the subsequent planet formation process.
The growth model that we 
adopted from \citet{2005A&A...436..933F} calculated the growth status
after one free-fall time. We will investigate this question further in
a forth-coming paper dealing with the growth and maximal sizes expected
for molecular cloud cores.

To sum up, we refer to the two goals formulated at the beginning of this
section:
(i) we have demonstrated that scattered light
images based on a density structure that agrees with the
current extinction data of L183 can reproduce the main
features of the observed Spitzer images. Some deviations arise from 
numerical artefacts or incompleteness of the extinction data and are
expected to be tackled with a better modeling; 
(ii) the presented images use a grain growth model, and indeed the
images are very sensitive to the threshold density where growth starts,
to the starting grain size, and to the growth efficiency. We find that
models without dust growth do not produce any measurable scattered light flux,
while a growth from grains with a size of about 0.05 $\mu$m starting
at an H$_2$ number density of 
$3\times 10^{10}$ m$^{-3}$ can reproduce the emission pattern seen
by Spitzer in the 3.6 and 4.5 $\mu$m filter.

\section{Summary} \label{conclusions}

In this paper, we have presented Spitzer maps of the dark cloud
complex L183 which reveal emission
at 3.6 and 4.5 $\mu$m and absorption at 8 $\mu$m near the
densest regions of the cloud (\Av\ $>15$). 
We argued that the emission cannot
be attributed to thermal black body emission of larger dust grains, they are too cold
in this environment.
The emission of stochastically heated particles should originate in
the layer where most of the interstellar radiation field is absorbed.
The morphology of this layer is visible in the scattered light at
optical wavelengths and traces the outer contours of the extinction map,
not the inner banana-shaped region of L183.
Moreover, existing PAH emission models show a strong increase in emission
from 4.5 to 5.8 $\mu$m, which is not seen in the Spitzer images.
Finally, a careful investigation of the 
density structure has confirmed that an excitation of stochastically
heated particles (PAHs and small grains) through holes in the clumpy
cloud is
unlikely, because of the optical depth for the interstellar radiation at
optical and UV wavelengths for almost all investigated
spatial structure models for L183
(see Appendix A). 
We therefore suggested that this emission is scattered interstellar 
radiation.

To test this hypothesis, we performed 3D radiative transfer calculations producing scattered light images of the central dense region where
emission is seen.
The density structures were derived from an automated fit of the
\Av~map for this region (from Paper I)  by a series of 3D Gaussian
number density functions. The missing information along the line-of-sight
was dealt with by creating a library of possible density structures
in agreement with the \Av~map.

When using simple silicate dust grains without growth, the produced
scattered light flux was not sufficient to reproduce the observed image.
This finding was independent of the
chosen spatial structure and the dust grain size.
The model included a realistic direction-dependent ISRF based on
the DIRBE data, which did not introduce new unknowns that could change
this view.

This led us to consider grain models that allow for grain growth
in the denser parts of the cloud. Assuming a simple growth law
along the directions of recent laboratory measurements, we investigated
the changes in the optical depth and the scattering direction for
photons propagating into regions with increasing density 
containing larger dust grains.

Within this model, we were able to reproduce the
main features of the Spitzer observation in the 3.6, 4.5, and 8
$\mu$m bands: 
(i) the emission arises mainly in the dense regions, 
(ii) the emission in the 4.5 $\mu$m band is weaker than in the 
     3.6 $\mu$m band and concentrated in the densest upper part of
     the banana-shaped central region,
(iii) the central region appears in absorption at 8 $\mu$m.
The flux is smaller by a factor of about 2 in the model images compared
to the 3.6 $\mu$m flux measured by Spitzer.
However,
various uncertainties enter the modeling, namely the uncertainties in
the \Av~map, the errors in the \Av~map fitting, the uncertainties 
in the extinction properties of the dust, the uncertainty in the
scattering phase function, the approximations in the dust model,
and the uncertainty in the ISRF.

The existence of a new MIR window to study the growth of grains
in the dense parts of cores is the result of an interplay of several
effects: At MIR wavelengths, the moderate optical depth allows
observing dense regions where growth is expected. The
MIR scattering efficiency of smaller grains is too low to detect them,
but larger grains scatter more efficiently and become observable again.

From our results we conclude that this is the reason why we see
emission in the 3.6 and 4.5 $\mu$m Spitzer
images of L183 as interstellar radiation scattered by dust
particles grown in the dense region with respect to
common ISM particles.
Therefore, the Spitzer observations of L183 
are the first {\em direct} evidence of dust growth in molecular
cloud cores.
In turn, we evaluate the evidence
from the {\em extinction} measurement at MIR and from FIR/mm observations
to be {\em indirect} as there is an ambiguity in explaining the 
data by either assuming larger grains or by modifying the density
and temperature structure used in the LoS integration.

In a future paper, we will present a 3D model of the dust based on a
revised \Av~map; and it will be strongly constrained by combining all
available data from mm to NIR.

\begin{acknowledgements}

This research has made use of NASA's Astrophysics Data System Abstract 
Service. 
We are thankful to 
Fran\c cois M\'enard, Nikolai Voshchinnikov, Thomas Henning, Cornelius
Dullemond, and 
Roy van Boeckel
for fruitful discussions and to Nicolas Flagey for his help with Spitzer data.

\end{acknowledgements}

\bibliographystyle{aa}
\bibliography{l183}

\begin{appendix}
  \section{Exploring the applicability of a PAH model:
              Three-dimensional structure variations and
              $\tau$-iso-surfaces} \label{structure}

From the \Av~map alone, it cannot be excluded that stellar emission is
approaching the inner parts of L183 to excite PAHs at the locations where
emission is seen in Spitzer images. 
Projection effects can hide holes which would allow such excitation, and
would not to be unlikely given the clumpy structure of the core.
Becausethe PAHs are excited by optical and UV photons, we consider
a wavelength of 0.55 $\mu$m as a conservative limit - if this radiation is
not able to reach the PAHs then radiation at shorter wavelengths will not
either. In this analysis, we define ``reaching'' as $\tau<1$.
We performed spatial structure analysis calculations to explore the 
cases for which PAH excitation by the ISRF is possible.
Varying the extent and position of the Gaussians along the line-of-sight
randomly 
we obtained 10000 spatial structures, each of it in agreement with the
L183 \Av~map.

In the following, we would like to examine which of these spatial structures
would allow for a PAH emission model.
To characterize each configuration, we calculated 
the minimal 3D distance of the 
$\tau_{0.55\mu m}=1$-surface to the central density maximum $D_m$ 
(as an indicator for density holes) and 
the minimal optical depth at 0.55 $\mu$m toward the
density maximum $\tau_m$ (as indicator for the heating
of the center by the ISRF).
\begin{figure}
\centering
\includegraphics[width=9cm]{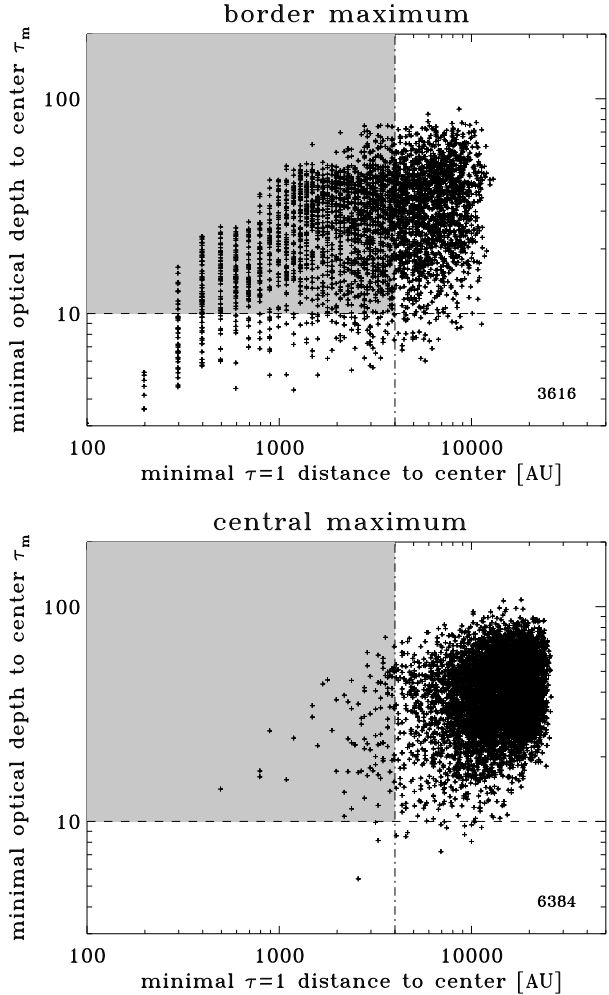}
\caption{
Minimal optical depth at 0.55 $\mu$m $\tau_m$ to the central density 
maximum as a function of the 3D distance of the
$\tau_{0.55\mu m}=1$-surface to the central density maximum.
The upper and lower plots show spatial structures for which the central density
maximum is located within and outside the inner 30000 AU of the 
spatial structure, respectively. The numbers in the lower right corner indicate
the number of spatial structures in this category.
        }
\label{tau1mindisttaumin}
\end{figure}
The results are plotted in 
Fig.~\ref{tau1mindisttaumin}, 
which relates both quantities. 

To distinguish the spatial structures with a central density
maximum close to the front or back parts of the considered density cube
(which is close to the low-density outer cloud part)
and those that have the center in the middle part, we
plotted those spatial structures in the upper (lower) panels for which the central density
maximum is located within (outside) the inner 30000 AU of the
spatial structure, respectively. In the following, we refer to these two cases
as {\em central} and {\em border} maximum spatial structures. Their numbers are
given in the bottom right corner of each plot.

In the 3.6 $\mu$m Spitzer image, emission is seen out to $D_E$=4000 AU
measured in the plane of sky from the central density maximum.
Assuming that the emission comes from the surface of a cylinder with this
radius, the ISRF has to enter the cloud down to this distance 
to excite PAHs.
The dash-dotted line indicates this
distance $D_E$ and 
the $\tau_{0.55\mu m}=1$-surface should be closer to 
the center or efficient PAH excitation will not be possible. As a conservative criterium, we assume that minimal
distance of the $\tau_{0.55\mu m}=1$-surface to the center should be less than $D_E$.

For most border maximum spatial structures and almost all central maximum
spatial structures (right of the dash-dotted line)
this is not the case. Exciting stellar optical light 
cannot reach the inner parts without being heavily extincted, so that
PAH
emission can be excluded as an explanation for the observed emission.
A few border maximum spatial structures reach low $\tau_m<10$ values.
The resulting temperature of the central density peak would be higher
than the 7-9 K found by \citet{2003A&A...406L..59P,2004A&A...417..605P}. 
Thus, spatial structures
below the dashed line at $\tau_m<10$ are also not consistent with the 
observational data.

We are therefore left with spatial structures in the upper left hand
corner of the
two plots in Fig.~\ref{tau1mindisttaumin} to be candidates for 
possible PAH emission models. For the upper plot, the central density
maximum can be located in the front part or in the back of the entire
spatial structure. If it is located in the back, the excited PAH emission has
to cross the entire spatial structure. 
Since $\tau_{3.6\mu{m}}$ is about an order of magnitude less
than $\tau_{0.5\mu{m}}$, the expected optical depth would still be 
between a few to ten, making the PAH emission produced in the front part
dominate the emission.

If it is located in the front, it will be illuminated well, hence
expected to be seen in emission. Only if there is more flux created
in the back by the other less dense clumps would it be seen in absorption.
This region between the other clumps and the foreground clump are
shielded well though, because to excite PAHs, radiation has to cross either the
other clumps or the foreground clump.

Limb brightening in an optically thick clump would produce a ring
of high surface brightness and a dip in the center. But such an external
dense clump should be seen in the optical images of L183 as a small region
with enhanced scattered light. But the CFHT I image only shows 
smooth scattered light on scales of the outer $\tau=1$ layer visible
in the \Av~map. Moreover, just a dip is seen but not a ring.

In view of these arguments, it appears unlikely that the central part
of L183 has a density maximum located near the edge of the
spatial structure.
The remaining few central maximum spatial structures in the lower plot of
Fig.~\ref{tau1mindisttaumin} left from the $D_m$ dash-dotted line
and with $\tau_m>10$ are the only spatial structures with holes arranged
in a way that a PAH emission model would work.

With 10000 random spatial structures, a parameter space of 200 parameters
is not well-sampled, but the main optical depth effects are dominated
by a few ten Gaussians. From this analysis, we conclude that 
a PAH model for the 3.6 $\mu$m emission cannot be excluded formally 
with the 
available data. From our sample of 3D spatial structures
consistent with the \Av~map, only a very small fraction would allow for
PAH emission near the central absorption peak, which therefore seems unlikely.

  \section{A possible star within L183} \label{star}

Beside the ISRF, another source of illumination
could be the presence of a 
nearby star, positioned at $\sim$3' away from the reference 
position. 
This star, visible in neither the blue POSS plates nor the red POSS I 
plate and hardly visible in the red POSS 2 plate, is bright in the 
infrared (Table \ref{magstar}), implying that the star is attenuated by
the cloud. 
We started by deriving a set of plausible reddening values 
(from the H and 
K' difference, we found \Av\ in the range 10 to 23 mag). For this 
range, no 
normal star (main sequence, giant, supergiant) could fit the data.

\begin{table}[htdp]
\caption{Star position and magnitudes}
\centering
\begin{tabular}{cccc}
\hline
            \noalign{\smallskip}
Position& R.A.&Dec\\
            \noalign{\smallskip}
\hline
            \noalign{\smallskip}
(J2000)&15$^h$54$^m$20.45$^s$ &-02$^o$54'07.30"\\
            \noalign{\smallskip}
\hline
\hline
            \noalign{\smallskip}
Band&Brightness&Error&Ref\\
($\mu$m)&mag&mag&\\
            \noalign{\smallskip}
\hline
            \noalign{\smallskip}
R&(19.7)$^{\mathrm{a}}$&&1\\
I&16.3&0.1&2\\
J&10.986&0.021&3\\
H&8.915&0.044&3\\
K&7.910&0.016&3\\
3.6&7.361&0.002&4\\
4.5&7.340&0.002&4\\
5.8&7.222&0.003&4\\
8&7.177&0.003&4\\
24&7.4&0.1&4\\
            \noalign{\smallskip}
\hline
\end{tabular}
\begin{list}{}{}
\item[$^{\mathrm{a}}$] USNO-B calibrations are done in the original E band plate, which differs somewhat from the standard R band \citep{2003AJ....125..984M}
\end{list}
References : (1) USNO-B \citep{2003AJ....125..984M}, (2) unpublished data from Paper I, (3) 2MASS \citep{CutriR.M.2000}, (4) this work.
\label{magstar}
\end{table}%

\begin{figure}
\centering
\includegraphics[width=9cm]{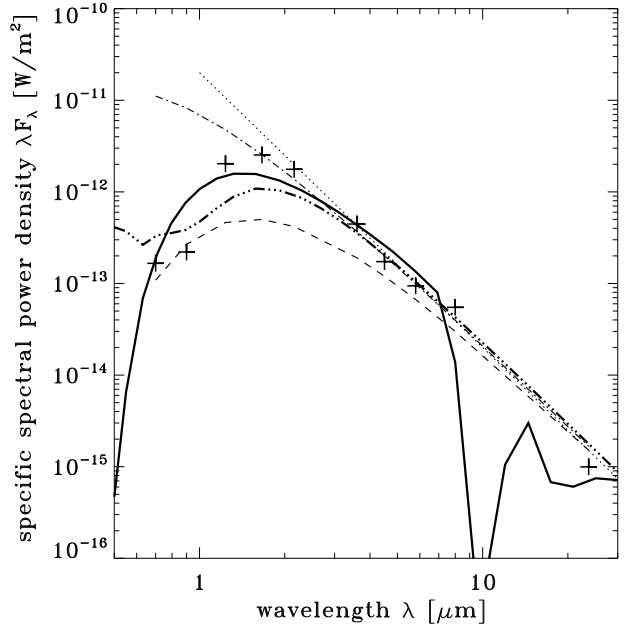}
\caption{
Specific spectral energy density as a function of the wavelength.
The observed data points are shown as thick crosses.
Three fits are displayed without extinction are displayed: a
spectrum in the Rayleigh-Jeans limit (dotted),
a Planck spectrum having the same maximum
as the observed SED (dashed), and 
a Planck spectrum that follows the data where the deviation from the
Rayleigh-Jeans spectrum occurs (dashed-dotted).
For the case of extincted stellar emission, two spectra are shown:
a Planck spectrum of a star being extincted by silicate grains 
(thick solid), and by carbon grains (dash-three dotted), respectively.
        }
\label{SED}
\end{figure}
We therefore applied a more detailed analysis, by applying an automated SED fitting procedure based on the simulated
annealing code \citep{1994A&A...287..493T} that calculates the
spectrum using an extincted Planck spectrum varying
the effective stellar surface
temperature $T$, the distance , 
the stellar radius $R$, and the column density of
extincting material for ISM Si and C grains.
Figure \ref{SED} shows the specific
spectral energy density $\lambda F_\lambda$
as a function of $\lambda$.
The data points are marked as thick crosses, no error bars have been
considered as the given errors are small (see Table \ref{magstar}).
Comparing with an unextincted 
spectrum in the Rayleigh-Jeans limit (dotted line), 
no indication of IR excess is visible, which would arise from the thermal
emission of warm dust in a nearby disk and/or shell.

We tried two ways to fit the spectrum with a stellar spectrum that 
undergoes no extinction. 
The dashed line gives a fit for a Planck spectrum that has the same maximum
as the observed SED ($T_*=2400$ K, $R_*>3\times10^6$ km, same distance
as L183).
Cool object spectra fail to reproduce
the sharp turnover around the maximum without extinction.
The dash-dotted lines shows the Planck spectrum without extinction that follows the data where the deviation from the
Rayleigh-Jeans spectrum occurs
($T_*=7000$ K, $R_*>2\times10^6$ km), but 
the flux decrease is too weak toward shorter wavelengths.
Hotter objects cannot reproduce the sharp turnover either.
We therefore concluded that extinction is needed to achieve better
fits.

Putting a layer of extincting dust in front, we used the SED optimizer
to investigate two cases: 
(i) silicate grains (solid thick line)
and (ii) carbon grains (solid dash-three dotted line)
\citep{1984ApJ...285...89D}. 
The thick solid line gives the fitted spectrum for silicate grains, the
 dash-three dotted line for carbon grains. 
Both fits failed to reproduce the SED with
good accuracy.
In these models, the temperature was around $T_*=7000$ K, and the
extinction was about $\tau=$5 to 9 depending on the dust model.
It would need a more sophisticated dust and/or stellar atmosphere model
to improve the fits.
Given the projected position in the extinction map, 
the optical depth from the \Av~map for the star
when sitting behind the core would be 10 to 15. This value is close to
the extinction derived from the SED fitting.
We conclude from these data that the star is not in front of the cloud. 

We calculated the flux ratio of a star consistent with the SED fits at 
various distances from L183 and the ISRF at 3.6 $\mu$m using DIRBE data. 
We find that the radiation field is dominated by the star up to about 
a distance of 2000 AU and then the stellar influence drops quickly due to 
the quadratical decrease with distance.
If the star were placed near L183 or even exactly at the same distance to Earth 
as the density maximum, it still would be located in the outer parts of 
L183 as is evident in Fig.~1, and we would only expect a very local 
(a few thousand AU) influence on the scattered light pattern, 
while the ISRF dominates for the entire banana-shaped region. 
The potential influence of the star being placed near L183 is
weakened further by its having to be near a dense region with grown 
particles to produce scattered light at 3.6 micron.
\end{appendix}

\end{document}